# RBF neural net based classifier for the AIRIX accelerator fault diagnosis

J.C. Ribes, G. Delaunay
LAM - Université de Reims Champagne-Ardenne - F51687 REIMS cedex2
J. Delvaux, E. Merle, M. Mouillet
CEA - PEM   F51490 Pontfaverger - France

*Abstract*

The AIRIX facility is a high current linear accelerator (2-3.5kA) used for flash-radiography at the CEA of Moronvilliers France. The general background of this study is the diagnosis and the predictive maintenance of AIRIX. We will present a tool for fault diagnosis and monitoring based on pattern recognition using artificial neural network. Parameters extracted from the signals recorded on each shot are used to define a vector to be classified. The principal component analysis permits us to select the most pertinent information and reduce the redundancy. A three layer Radial Basis Function (RBF) neural network is used to classify the states of the accelerator. We initialize the network by applying an unsupervised fuzzy technique to the training base. This allows us to determine the number of clusters and real classes, which define the number of cells on the hidden and output layers of the network. The weights between the hidden and the output layers, realising the non-convex union of the clusters, are determined by a least square method. Membership and ambiguity rejection enable the network to learn unknown failures, and to monitor accelerator operations to predict future failures. We will present the first results obtained on the injector.

## 1. INTRODUCTION

The AIRIX induction accelerator [1] is used for flash radiography . The single shot functioning needs to be optimal at the desired time. We try to define a supervision method for fault diagnosis for the AIRIX facility. The goal is to garanty optimal functioning and to search precursors for the known failures in developping predictive maintenance . We are interessed to diagnosis using pattern recognition with neural network, and particularly RBF nets. RBF neural nets are employed for supervised classification. We propose an original strategy to create RBF nets for unsupervised classification.

First we present the experimental context and the goal of the study. In the following paragraphs, we develop the strategy used for defining an adaptative supervised classifier with unsupervised data. Finally, we expose the first results obtained for the injector based on data from the year 1999 to the first trimester 2000.

## 2. EXPERIMENTAL CONTEXT

The AIRIX facility consists in an injector, 32 high voltage generators and 64 induction cells. The single shot functioning imposes to obtain the best performances at a given time. We try to develop a predictive maintenance based on signal processing and pattern recognition. Firstly this application is an help for users, principaly to have a most precise diagnosis and an automatic quick view of the state of the accelerator during the experiments. Secondly the results of automatic recognition could be an help to detect functioning drift to plan maintenance operations.

About 300 signals are recorded during each experiment. The complexity of the installation imposes to decompose the different module into sub-systems to realize a precise diagnosis. Signal processing permits to caracterize electrical signals of the machine. Those parameters define the state of functioning as a vector to be classified. We focus on the injector to develop a prototype of supervisor. The injector [2], which creates the electron beam, can be separated in three main modules : the prime power, the laser commanded spark gaz and the vacuum diode. The electron beam (4MeV,2kA,60ns) is created by applying on the cathode of the diode a pulse of 4MV. The prime power, which furnishes the primary high voltage, is caracterized by 8 parametres. For the sparks, which allow the discharge of the prime power into the diode along 3 transfert lines, 15 parametres are used. The firing of the 4 sparks must be realized simultaneously with approximativly the same current level. Finally the vacuum diode, in which the electrons are emited is diagnosed by 18 features. Those parameters are characteristical values, such as rise time, half height, temporal position of the different peaks and frequency values. The collected data are used to define a supervised training base for an automatic classification of failures.

The first step (part3) is to analyse the unsupervised data to define a training set and the different existing states. The second step (part 4) consists in classifying the data with an artificial supervised neural network.

## 3. UNSUPERVISED DATA ANALYSIS

The fuzzy-c-means algorithm [3] allows to cluster data by minimizing the following fuzzy criteria.

$$J = \sum_{i=1}^{m} \sum_{j=1}^{n} (u_{ij})^p \cdot d^2(x_j, g_i)$$

m is the number of classes and n the number of points. d is the euclidian distance and $u_{ij}$ the fuzzy membership of data $x_j$ to the class i. Each cluster is defined by a center or prototype note gi. p is the fuzzyfication of the criteria and is generally taken to 2.

The algorithm is composed of the following steps.
- *initialise m, g and U.*
- *Start with an initial partition*
- *Determine the center $g_i$ with*

$$g_i = \sum_{j=1}^{n}(u_{ij})^p \cdot x_k \Big/ \sum_{j=1}^{n}(u_{ij})^p$$

- *update U with* $u_{ki} = 1 \Big/ \sum_{j=1}^{m}\left(\frac{\|x_i - g_k\|^2}{\|x_i - g_j\|^2}\right)^{\frac{1}{p-1}}$

- *Cluster with the new centers*

The algorithm is iterated until the stability of the prototypes. The principal drawback of this technic is that we impose the number of class. The compacity criteria allows to determined the sufficant number of protoype to cluster the data. We use it for a multiprototype approach of the clustering. A class is caracterized by multiple centers. The number of real class in the data is determined by cutting a hierarchical tree calculate on the centers. We used the minimization of the function K as a criteria [4] to resolve the problem of calculating the level of cutting.

$$K = \left|1 - \frac{co_{gl}}{co_{moy}}\right|$$

with $co_{gl} = \frac{1}{n}\sum_{j=1}^{m}\sum_{i=1}^{n}[u_{ij} \cdot d(x_j, g_i)]^2$

$$co_{moy} = \frac{1}{m}\sum_{j=1}^{m}\frac{\sum_{i=1}^{n}[u_{ij} \cdot d(x_j, g_i)]^2}{\sum_{i=1}^{n}u_{ij}}$$

*co* is the compacity in a multiprotoype approach.
This method has permited to define the different classes and a supervised training data set. The defined classes have been succesfully physically identified.

## 4. RBF NEURAL NET CLASSIFIER

They are constructed with 3 layers (fig 1). The hidden neurons use a non linear radial basis function activation [5]. We use a classical gaussian form for this function.

$$\varphi(x_j) = \exp\left\{-\frac{d^2(x_j, g_i)}{2\sigma^2}\right\}$$ where d is the euclidian distance between an observation and the considered neuron and σ is the size of the gaussian. The weigths between the hidden and the output layer are adjust during the training step.

The structure of the neural network is presented on the next figure.

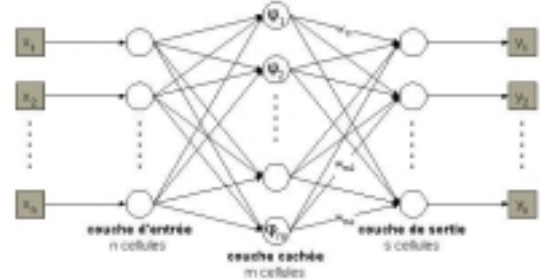

Figure 1 : RBF neural network

The classification is realized on the maximum membership principle . A data is affected to the class with the maximal output, determined by :

$$y_k(x_j) = \sum_{i=1}^{m} w_{ik} \varphi(x_j)$$ avec k=1,2,…s.

The most important problem is to define the number of cells on the hidden layer.

We propose to use the results of the precedent section to define the number of neural cells. The centers, which define the clustering can be interpreted as the hidden neurons and the cut of the tree defined the number of output neural cells.

Each hidden neuron is now seen as a fuzzy set with a gaussian membership function of size σ.

The parametre σ is calculated with the following formula.

$$\sigma = \frac{1}{2}\min(d(g_{kj} - g_{li}))$$

$g_{kj}$ is the hidden neuron i belonging to the class k and $g_{li}$ the neuron i of a class l with k $k \neq l$.

Finally the network is trained with the known data defined at the paragraph 3. The training consists in resolving a linear system with the least square algorithm. The variables to determine are the weight of the last layer and the output is 1 if the data belong to the class and 0 otherwise.

With this last step, we dispose of a model able to classify the data initialy unsupervised, without the knowledge of the number and the membership of data to class. In case of unknown state the reject options are used. They allow to detect observations wich can't be reliably classified. An observation too far from the others is membership rejected. If an observation is too close from different classes, it is ambiguity rejected The first reject is a way to detect knew state for adaptation of the classifier. The second reject permits to detect the evolution of the functioning from one class to another.

## 5. RESULTS

We have defined an observation space for the three subsystems of the injector. The unsupervised analysis of the data permits to detect and to learn differents state of the injector's functioning. The AIRIX facility is operational since the beginning of the year 2000. We know only few failures, but we are able to caracterize the different levels of functioning. The injector is used with three level corresponding to the voltage applyed to the vacuum diode (initial energy of the beam), 2.3MeV 3MeV and 4 MeV (nominal mode).

Each sub system contains at the initialization 3 classes, one for each level, the other classes corresponding to the known failures . The following figures present the space of decision for the injector's subsystem (fig 2, 3 4). We note that the differents classes are well discrimated and allow a good decision making.

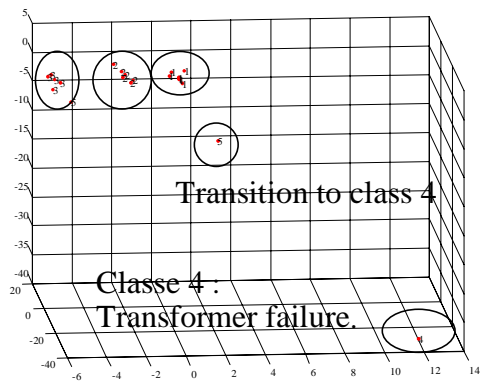

Figure 2 : prime power

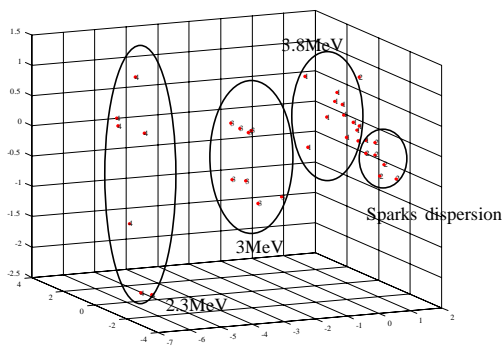

Figure 3 : spark gaz

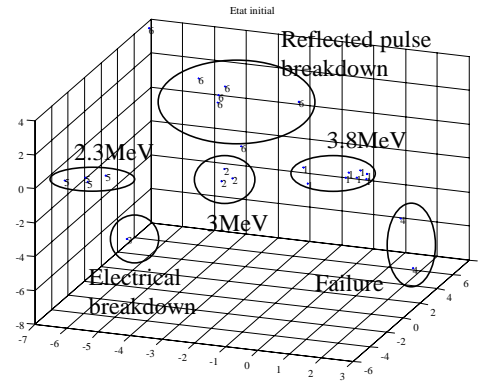

Figure 4 : vacuum diode

The most current state of functioning and the principal failures (electrical breakdown on the diode's pannel, dispersion on the spark functioning) have been learned by the classifier. We have obtain the following results of good recognition with our prototype.

Table 1 : results of classification

| System | Rejected | error | succes |
|--------|----------|-------|--------|
| Prime  | 43       | 10    | 380    |
| spark  | 32       | 5     | 402    |
| diode  | 44       | 0     | 390    |

The number of rejected point could be ameliorated with an adaptation of the classifier to new states. Mistake in decision have been identified as a confusion between levels of energy and never as the non detection of a failure.

## 6. CONCLUSION

A prototype of classifier with RBF neural net is now built. The good properties of the RBF neural net allow a good decision making, completetd with the reject options, which improve the reliability of the results. This strategy permits a good diagnosis of the installation. We must extend the results to the whole accelerator and use them for the maintenance strategy.